\documentclass[letter,longauth]{aa}  
\usepackage{graphicx}
\usepackage{natbib}
\bibpunct{(}{)}{;}{a}{}{,} 
\usepackage[varg]{txfonts}
\usepackage{hyperref}
\def\ms{\hbox{\,m\,s$^{-1}$}}         
\def\m2s2{\hbox{\,m$^{2}$\,s$^{-2}$}} 
\def\sini{\hbox{sin\,$i$}}      

\def\Mjup{\hbox{$\mathrm{M}_{\rm Jup}$}}
\def\Rjup{\hbox{$\mathrm{R}_{\rm Jup}$}}

\begin{document}

   \title{The GAPS Programme with HARPS-N@TNG}

   \subtitle{IX. The multi-planet system KELT-6: Detection of the planet KELT-6\,c and measurement of the Rossiter-McLaughlin effect for KELT-6\,b  \thanks{Based on observations made with (\textit{i}) the HARPS-N spectrograph on the Italian Telescopio Nazionale Galileo (TNG), operated on the island of La Palma by the INAF - Fundacion Galileo Galilei (Spanish Observatory of Roque de los Muchachos of the IAC); (\textit{ii}) the Tillinghast Reflector Echelle Spectrograph (TRES) on the 1.5-meter Tillinghast telescope, located at the Smithsonian Astrophysical Observatory's Fred L. Whipple Observatory on Mt. Hopkins in Arizona; (\textit{iii}) the IAC-80 telescope at the Teide Observatory (Instituto de Astrof{\'\i}sica de Canarias, IAC). Tables 2 and 3 are made available on-line at the CDS (http://cdsweb.u-strasbg.fr/).}}

\author{
M. Damasso     \inst{1},
M. Esposito    \inst{2,3},
V. Nascimbeni  \inst{4,5},
S. Desidera    \inst{4},
A.S. Bonomo    \inst{1},
A. Bieryla         \inst{6},
L. Malavolta   \inst{4,5},
K. Biazzo      \inst{7},
A. Sozzetti    \inst{1},
E. Covino      \inst{8},
D. W. Latham            \inst{6},
D. Gandolfi \inst{9,10}
M. Rainer      \inst{11},
C. Petrovich   \inst{12},
K. A. Collins  \inst{13,14},
C. Boccato \inst{4},
R.U. Claudi    \inst{4},
R. Cosentino   \inst{7,15},
R. Gratton      \inst{4},
A.F. Lanza     \inst{7},
A. Maggio          \inst{16},
G. Micela      \inst{16},
E. Molinari    \inst{15,17},
I. Pagano      \inst{7},
G. Piotto      \inst{4,5},
E. Poretti     \inst{11},
R. Smareglia       \inst{18},
L. Di Fabrizio \inst{15},
P. Giacobbe     \inst{1},
M. Gomez-Jimenez \inst{2,3},
S. Murabito \inst{2,3},
M. Molinaro \inst{18},
L. Affer       \inst{16},
M. Barbieri  \inst{19},
L. R. Bedin  \inst{4},
S. Benatti     \inst{4},
F. Borsa     \inst{9},
J. Maldonado   \inst{16},
L. Mancini     \inst{20,1},
G. Scandariato \inst{7},
J. Southworth  \inst{21},
R. Zanmar Sanchez \inst{7}
}

\institute{ 
INAF -- Osservatorio Astrofisico di Torino, Via Osservatorio 20, I-10025, Pino Torinese, Italy.\\
E-mail: \texttt{damasso@oato.inaf.it}
\and Instituto de Astrof{\'i}sica de Canarias, C/Via L{\'a}ctea S/N, E-38200 La Laguna, Tenerife, Spain
\and Departamento de Astrof{\'i}sica, Universidad de La Laguna,  E-38205 La Laguna, Tenerife, Spain
\and INAF -- Osservatorio Astronomico di Padova,  Vicolo dell'Osservatorio 5, I-35122, Padova, Italy
\and Dip. di Fisica e Astronomia Galileo Galilei -- Universit\`a di Padova, Vicolo dell'Osservatorio 2, I-35122, Padova, Italy
\and Harvard-Smithsonian Center for Astrophysics, Cambridge, MA 02138 USA
\and INAF -- Osservatorio Astrofisico di Catania, Via S.Sofia 78, I-95123, Catania, Italy
\and INAF -- Osservatorio Astronomico di Capodimonte, Salita Moiariello 16, I-80131, Napoli, Italy
\and Dipartimento di Fisica, Università di Torino, via P. Giuria 1, 10125 Torino, Italy
\and Landessternwarte K\"onigstuhl, Zentrum f\"ur Astronomie der Universit\"at Heidelberg, K\"onigstuhl 12, D-69117 Heidelberg, Germany
\and INAF -- Osservatorio Astronomico di Brera, Via E. Bianchi 46, I-23807 Merate (LC), Italy
\and Department of Astrophysical Sciences, Princeton University, Ivy Lane, Princeton, NJ 08544, USA
\and Department of Physics and Astronomy, Vanderbilt University, Nashville, TN 37235, USA
\and Department of Physics and Astronomy, University of Louisville, Louisville, KY 40292, USA
\and Fundaci\'on Galileo Galilei - INAF, Rambla Jos\'e Ana Fernandez P\'erez 7, E-38712 Bre\~na Baja, TF - Spain
\and INAF -- Osservatorio Astronomico di Palermo, Piazza del Parlamento, 1, I-90134, Palermo, Italy
\and INAF - IASF Milano, via Bassini 15, I-20133 Milano, Italy
\and INAF -- Osservatorio Astronomico di Trieste, via Tiepolo 11, I-34143 Trieste, Italy
\and Universidad de Atacama, Departamento de Fisica, Copayapu 485, Copiapo, Chile
\and Max Planck Institute for Astronomy, K\"{o}nigstuhl 17, 69117 -- Heidelberg, Germany 
\and Astrophysics Group, Keele University, Staffordshire, ST5 5BG, UK
}

   \date{}

 
  \abstract
   {}
   {For more than 1.5 years we spectroscopically monitored the star KELT-6 (BD+31\,2447), which is known to host the transiting hot-Saturn KELT-6\,b, because a previously observed long-term trend in radial velocity time series suggested that there is an outer companion.}
   {We collected a total of 93 new spectra with the HARPS-N and TRES spectrographs. A spectroscopic transit of KELT-6\,b was observed with HARPS-N, and simultaneous photometry was obtained with the IAC-80 telescope.}
   {We proved the existence of an outer planet with a mininum mass $M_{\rm p}\sini$=3.71$\pm$0.21 $M_{\rm Jup}$ and a moderately eccentric orbit ($e=0.21_{-0.036}^{+0.039}$) of period P$\sim$3.5 years. We improved the orbital solution of KELT-6\,b and obtained the first measurement of the Rossiter-McLaughlin effect, showing that the planet has a likely circular, prograde, and slightly misaligned orbit with a projected spin-orbit angle of $\lambda$=$-$36$\pm$11 degrees. We improved the KELT-6\,b transit ephemeris from photometry and provide new measurements of the stellar parameters. KELT-6 appears as an interesting case for studying the formation and evolution of multi-planet systems.} 
  {}


   \authorrunning{M. Damasso et al.}
   \titlerunning{GAPS IX. The KELT-6 multi-planet system}

   \maketitle
%

\section{Introduction}

Analysing stellar radial velocity (RV) time series is an effective method of detecting and characterizing distant planets with orbital periods of a few years, despite the large observing time span required. To look for such companions, particularly interesting targets are stars with transiting planets, because they are an ideal laboratory for studying the architecture of multi-planet systems. In fact, the orbital geometry of the transiting planet can be described through the Rossiter-McLaughlin (RM) effect (e.g. \citealt{ohta05}) by measuring the projected spin-orbit angle and spotting the direction of the motion with respect to that of the stellar rotation. Up to now, long-term trends have been observed in the RVs of a large sample of stars with and without evidence of turnover (e.g. \citealt{knutson14}). Among them is KELT-6, a late F-type, metal-poor star (V=10.3 mag) hosting the transiting Saturn-mass planet KELT-6\,b discovered by the KELT-North survey (\citealt{collins14}, hereafter Co14). Co14 were also able to observe an unexplained residual trend in the RVs over a limited time span of 475 days. To understand the cause of this acceleration, a spectroscopic follow-up was carried out in the framework of the Global Architectures of Planetary Systems (GAPS) project\footnote{\begin{tiny}
http://www.oact.inaf.it/exoit/EXO-IT/Projects/Entries/2011/12/27$\_$GAPS.html 
\end{tiny}}, using the HARPS-N spectrograph (resolving power $R=115\,000$; \citealt{cosentino12}). We also collected new RV data with the Tillinghast Reflector Echelle Spectrograph (TRES) spectrograph ($R=44\,000$; \citealt{furesz08}), extending the total observing time span to 1178 days. 

Together with new photometric measurements, we present results that noticeably extend knowledge about the KELT-6 system.

\section{Observations and data reduction methods}
\label{sec:datar}

We collected 71 HARPS-N  spectra (exposure 900~s, typical signal-to-noise per pixel S/N$\sim$60 at 5500 $\AA$) between 2014 February 9 and 2015 July 3, 31 of which were obtained on 2015 April 11 during a transit of KELT-6\,b and used to study the RM effect. The Th-Ar simultaneous calibration was not used to avoid contamination by the lamp lines. The spectra and the RV measurements were reduced using the latest version (Nov. 2013) of the HARPS-N instrument Data Reduction Software pipeline and applying a G2 mask. The measurement of the RVs is based on the weighted cross-correlation function (CCF) method \citep{baranne96,pepe02}.

With TRES we collected 22 spectra between 2013 December 13 and 2015 May 27. They were extracted following the procedures described by \cite{buchhave10}. The relative RVs were derived by cross-correlating the spectra against the highest S/R spectrum in the wavelength range 4050$-$ 5650 $\AA$.

Simultaneously with the RM effect measurements gathered with HARPS-N, we collected the transit light curve with the IAC-80 0.82-m telescope. Data were taken from 21:18 UT to 5:32 UT, using the CAMELOT camera (E2V 2k$\times$2k CCD; pixel scale 0.304$^{\prime\prime}$; field of view $10.4^{\prime}\times 10.4^{\prime}$) and through a standard Bessell $R$ filter. The point spread function (PSF) was intentionally defocused to a radius of $\sim20$ physical pixels to minimize flat-field residual errors and avoid detector saturation. The exposure time was set to 90~s, resulting in a net cadence of $\sim$115~s when considering the overheads. Science frames were bias- and flat-field-corrected by standard procedures. Photometric measurements were made with the STAR$\-SKY$ pipeline \citep{nascimbeni11,nascimbeni13}. STAR$\-SKY$ delivered the best differential light curve of KELT-6 (i.e. the one with the least scatter) using a set of four stable comparison stars (UCAC4 604-049448, UCAC4 604-049449, UCAC4 604-049450, and UCAC4 604-049454).
Unfortunately, our observations were plagued by technical problems that delayed the start time of the observations, and the telescope pointing was re-adjusted, causing an offset between the first and the second halves of the light curve, which was included as a parameter in the transit model. 

\section{Stellar parameters}
\label{sec:stellarpar}

The photospheric parameters were derived with different methods from the co-added HARPS-N spectrum obtained from the out-of-transit observations (S/R$\sim$380 per pixel at 5500 $\AA$). We used the LTE code MOOG \citep{Sneden:1973el}, along with atmospheric models \citep{Kurucz:1992ab,Castelli:2004ti} and iron equivalent widths (EW). Two analyses were independently performed with the main differences being the list of iron lines and the technique used to measure EWs. In one case we used ARESv2 \citep{Sousa:2015ew} with the automatic continuum placement set-up and the line list from \cite{Sousa:2011gr}, adapted to the 2014 version of MOOG \citep{Dumusque:2014kp}. In the other case, the EWs of the line list from \cite{Biazzo:2012aa} were measured by hand using the IRAF task SPLOT, and the iron abundance was determined using the 2013 version of MOOG (see \citealt{damasso15}). In both cases, the analysis was performed differentially with respect to the Sun spectrum. We also did a third independent analysis based on the method described in \cite{gandolfi15}. We fitted the HARPS-N spectrum to a grid of theoretical models from \cite{Castelli:2004ti}, using spectral features that are sensitive to different photospheric parameters.

All the analyses gave consistent results, so we calculated their weighted averages and adopted the average of the individual uncertainties as errors. Table \ref{table:stellarparam} summarizes our results. We notice that our best-fit value for $T_{\rm eff}$ is 170 K higher than value adopted by Co14, which was obtained from spectral synthesis modelling with Spectroscopy Made Easy (SME; \citealt{valenti96}) using HIRES spectra. This difference could arise from the different analysis techniques rather than from the properties of the spectra, by noting that a well known bias exists in some versions of SME \citep{Torres:2012sp}, which was removed in more recent versions \citep{brewer15}.  

The stellar mass, radius, and age were determined by comparing our measured $T_{\rm eff}$, $\log g$, and [Fe/H] with the Yonsei-Yale evolutionary tracks \citep{demarque04} through the $\chi^{2}$ statistics \citep{santerne11}. Results are listed in Table \ref{table:stellarparam}. The adopted errors include an extra 5$\%$ in mass and 3$\%$ in radius added in quadrature to the formal errors to take systematic uncertainties into account in the stellar models \citep{southworth11}. We also employed the stellar density \textit{$\rho_{\rm \star}$} derived by Co14 as a proxy of the stellar luminosity instead of $\log g$ (e.g. \citealt{sozzetti07}). The results are fully consistent with the previous findings and not more precise. We thus adopted the parameters obtained using $\log g$, finding that KELT-6 appears to be slightly less evolved than stated by Co14. 
      

 \begin{table}
  \scriptsize
    \caption{Stellar parameters for the star KELT-6 derived from the analysis of the HARPS-N spectra and from stellar evolutionary tracks. Estimates from the KELT-6\,b discovery paper are also listed.}
    \label{table:stellarparam}
        \centering
        \begin{tabular}{cccc}
         \hline
         \noalign{\smallskip}
        Parameter         & This work & Co14 & Note \\  
         \noalign{\smallskip}
         \hline
         \noalign{\smallskip}
     T$_{eff}$ [K] & 6272$\pm$61 & 6102$\pm$43 & \\ [.5pt]
     $\log g$ [cgs] & 4.12$\pm$0.07  & 4.074$^{+0.045}_{-0.070}$ & \\ [.5pt]
     [Fe/H] [dex]  & -0.27$\pm$0.06  & -0.281$^{+0.039}_{-0.038}$ & \\ [.5pt]
     Microturb. $\xi$ [km s$^{-1}$]  & 1.49$\pm$0.1  & 0.85 (fixed) & \\ [0.5pt]
     \textit{V}sin\textit{I$_{\rm \star}$} [km s$^{-1}$] & 4.53$\pm$0.26 & 5.0$\pm$0.5 & (1) \\ [.5pt]
     $\mu$ linear law limb-darkening coeff. &  0.48$\pm$0.14 &  & (1) \\ [.5pt]
         Mass [M$_\odot$]  &  1.126$\pm$0.058  &  1.085$^{+0.043}_{-0.040}$  & (2) \\ [.5pt]
           Radius [R$_{\odot}$]   &  1.529$^{+0.143}_{-0.137}$      & 1.580$^{+0.160}_{-0.094}$  & (2)  \\ [.5pt]
           Age [Gyr]            &   4.90$^{+0.66}_{-0.46}$     & 6.1$\pm$0.2               & (2)  \\ [.5pt]
           Luminosity [L$_{\odot}$] &   3.24$\pm$0.62 & 3.11$^{+0.68}_{-0.39}$ & (2) \\ [.5pt] 
           Density [g cm$^{-3}$] & 0.44$^{+0.15}_{-0.10}$ & 0.387$^{+0.068}_{-0.088}$ & \\                      
          \noalign{\smallskip}
      \hline
        \end{tabular}
        \tablefoot{\scriptsize
         (1) Derived from HARPS-N spectra (RM effect).
         (2) Matching \textit{T}$_{\rm eff}$, [Fe/H], and $\log g$ to the Yonsei-Yale evolutionary tracks. 
         } 
    \end{table}


\section{Improved transit ephemeris for KELT-6b}
\label{sec:transit}

\begin{figure}
 \centering
 \includegraphics[width=7.5cm]{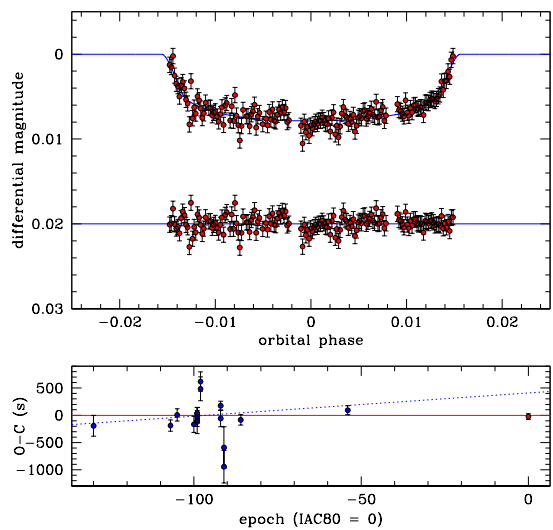}
 \caption{\emph{Upper panel:} KELT-6\,b transit light curve
(2014 Apr 11, IAC-80 telescope), and residuals from the best fit model (which is plotted with
a solid blue line). The root mean square of the residuals is 0.9 mmag. \emph{Lower panel:} $O-C$
diagram including the highest-quality measurements from C014 plus one unpublished measurement from the KELT team (blue points on the left side) and from our IAC-80 observation (red point). The $O-C$ diagram is folded on our newly determined ephemeris (solid red line), while the Co14 ephemeris is plotted with a dotted blue line.}
 \label{fig:lc}
 \end{figure}

The light curve observed with the IAC-80 telescope is shown in the upper panel of Fig.~\ref{fig:lc}. It has an average photometric scatter of 0.9 mmag on a 115~s timescale.

Since we could not measure the out-of-transit flux of KELT-6 to properly normalize the off-transit level of our light curve, a full, detailed modelling to extract all the orbital parameters\footnote{The fractional radii $R_\star/a$ and $R_\mathrm{p}/a$, with $R_\star$ being the star radius, $R_\mathrm{p}$ the planet radius, and $a$ the planet semi-major axis; the orbital inclination $i$; the orbital period $P$; the time of central transit $T_0$.} of KELT-6\,b was hampered by the unsolvable degeneracy between the normalization level and the other transit parameters. Therefore we chose to fix $R_\star/a$, $R_\mathrm{p}/a$, and $i$ to the best-fit values published by Co14, limiting the number of free parameters to four\footnote{i.e. $T_0$; the linear term $u_1$ of a quadratic limb darkening law, with the quadratic term interpolated from the \cite{claret11} tables; the off-transit normalization level; an existing zero-point offset between the first and the second halves of the light curve.}. The fit was performed with the code JKTEBOP v34 \citep{southworth08}, and the associated errors were derived through $10\,000$ classical Monte Carlo iterations. 

From the fit we could only determine the time of central transit $T_{\rm c}$ of KELT-6\,b. We then derived an improved ephemeris ($T_{\rm c}$ and orbital period $P$) from the observed-minus-calculated diagram (lower panel of Fig.~\ref{fig:lc}), by fitting both our measurement of $T_{\rm c}$ and the high-quality (``primary'') points from Co14 through a weighted least-squares procedure, as well as an additional ephemeris obtained by the KELT team from an unpublished transit that occurred on 2014 February 12. The results are shown in Table \ref{table:orbitalparam}, with the reference epoch set on the most recent transit.

\section{Analysis of the radial velocities}
\label{sec:radvelanalysis}
\subsection{KELT-6c comes out}
        
\begin{figure}
   \centering
   \includegraphics[width=\columnwidth]{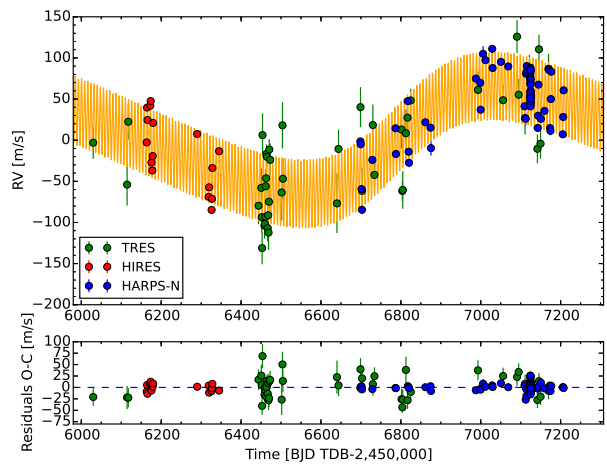} 
   \caption{\emph{Upper panel:} Radial velocity time series of KELT-6. The 1-$\sigma$ errorbars are over-plotted to each data point. When they are not visible, this means they are smaller than the symbol size. Over-plotted is our best two-planet Keplerian model (orange line), calculated from the best-fit orbital parameters of Table \ref{table:orbitalparam}. \emph{Lower panel:} Residuals of the best-fit two-planet model.}
   \label{fig:rvfit}
    \end{figure}

   The HARPS-N and TRES RVs collected for this work are listed in Tables \ref{table:radvelharps} and \ref{table:radveltres}, respectively.  In our sample, we also included the RVs used in Co14 and collected by TRES and the High Resolution Echelle Spectrometer (HIRES) at the Keck telescope, while we excluded the first 26 HARPS-N measurements taken during the night of the KELT-6\,b transit.      
  The RV time series shows a long-term modulation with a clear turnover and a semi-amplitude higher than that of the short-period signal due to KELT-6\,b (Fig. \ref{fig:rvfit}). KELT-6 appears to be a quiet star (<log(R$^{\prime}_{\rm HK}$)>=$-$4.992, $\sigma_{\rm \log(R^{\prime}_{\rm HK})}$=0.021), without evidence in the log(R$^{\prime}_{\rm HK}$) data for a modulation ascribable to an activity cycle (Fig. \ref{fig:rvcorr}, upper plot). Besides the log(R$^{\prime}_{\rm HK}$) index, we calculated the CCF bisector span (BIS) from the HARPS-N spectra, and we found that no significant correlations exist over the timespan of the HARPS-N measurements between these datasets and the RVs with the orbital solution of KELT-6\,b removed (Fig. \ref{fig:rvcorr}, middle and lower plots). This evidence supports the hypothesis that the observed RV long-term variations are not due to the stellar activity but to an outer companion, which we name KELT-6\,c and for which the data cover almost one complete orbit. This scenario is further strengthened by looking, for example, at the empirical calibrations of \cite{santos2000} (see Eq. 2 and Fig. 6 therein) that, for a star like KELT-6, predict a RV dispersion due to activity-related phenomena at the level of $\sim$10 m s$^{-1}$, while our measured semi-amplitude is $\sim$3-5 times higher ($K\sim$66 m s$^{-1}$) once the
orbital solution of KELT-6\,b is removed. Orbital parameters and uncertainties for the KELT planets were determined with a Bayesian differential evolution Markov chain Monte Carlo analysis (e.g. \citealt{desidera14}). We adopted a two-planet Keplerian model with sixteen free parameters\footnote{The central transit
epoch $T_0$, the orbital period $P$, the RV semi-amplitude $K$, $\sqrt{e}\,\cos\omega$, and $\sqrt{e}\,\sin\omega$ of  
both planets KELT-6\,b and KELT-6\,c ($e$ and $\omega$ being the eccentricity and the argument of periastron); a jitter term and a RV offset for each of the different datasets.}, assuming for KELT-6\,b Gaussian priors on $T_0$ and $P$ based on the new ephemeris. Fitted and derived parameters are listed in Table \ref{table:orbitalparam}. The best-fit model is shown in Fig. \ref{fig:rvfit}, and the residuals show a dispersion of $\sim$17 m s$^{-1}$, which reduces to 9.5 m s$^{-1}$ when considering only the more precise HIRES and HARPS-N measurements. This fact agrees nicely with the level of the activity-related jitter predicted by \cite{santos2000}. 
     
     KELT-6\,c appears to be massive ($M_{\rm p}\sini$=3.71$\pm$0.21 $M_{\rm Jup}$) and moves on a moderately eccentric orbit (significant at the $\sim$5$\sigma$ level) with period $P\sim$3.5 years. Residuals of the two-planet model do not reveal evidence of any trend of having a third companion (Fig. \ref{fig:rvfit}).     
     Our results suggest that KELT-6\,b has a likely circular orbit, with the eccentricity $e=0.029_{-0.013}^{+0.016}$  compatible with zero within $\sim$2.45$\sigma$ \citep{lucy71}. A reanalysis of the data done by the KELT team corrected the eccentricity to the value $e$=0.058$\pm$0.034, which is in accordance with our measurement.

\onlfig{
   \begin{figure}
   \centering
   \includegraphics[width=\columnwidth]{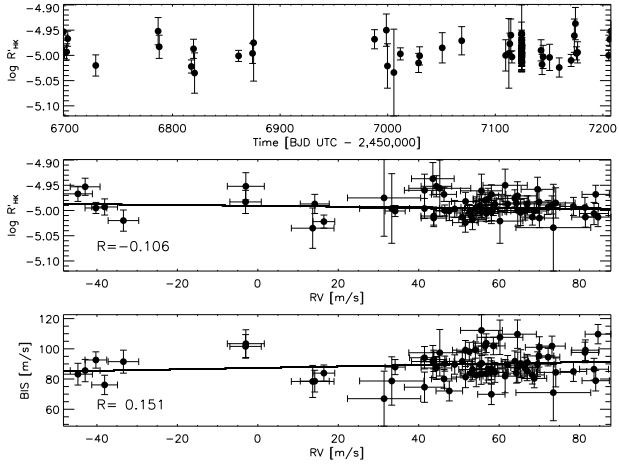} 
   \caption{ (\emph{Upper panel}). Time series of the log(R$^{\prime}_{\rm HK}$) chromospheric index as measured from HARPS-N spectra. (\emph{Middle and lower panels}). Correlation analysis between the RV residuals, obtained by removing the Keplerian signal due to KELT-6\,b from the original HARPS-N dataset, and two indicators of stellar activity derived from the HARPS-N spectra: the log(R$^{\prime}_{\rm HK}$) chromospheric index and the CCF bisector span. Over-plotted are the values of Spearman's rank correlation coefficients, which show the absence of significant correlations and support the hypothesis that the origin of the observed RV long-term variations is Keplerian.}
   \label{fig:rvcorr}
    \end{figure} 
    }

\onltab{
    \begin{table}
    \caption{Sample list of all the radial velocities (RV) measured for KELT-6 with the HARPS-N spectrograph in the framework of the GAPS programme. Also indicated are the bisector velocity span (BIS) and the activity index $\log$(\textit{R$^{\prime}_{\rm HK}$}). The BIS are those derived by the data reduction pipeline of HARPS-N with uncertainties assumed to be twice those on the radial velocities. 
}
    \label{table:radvelharps}
        \centering
        \scriptsize
        \begin{tabular}{cccccc}
        \hline
        \noalign{\smallskip}
        \noalign{\smallskip}
        \hline
    \noalign{\smallskip}
    Time  & RV & RV err. & BIS & $\log$(\textit{R$^{\prime}_{\rm HK}$}) & $\log$(\textit{R$^{\prime}_{\rm HK}$}) err. \\  
        (BJD$_{UTC}$ - 2,450,000) & (m s$^{-1}$) & (m s$^{-1}$) & (m s$^{-1}$) & &  \\ 
 6698.624617 & 1127.6  &  3.7 & 85.6 & -4.953 &  0.017 \\
 6699.598118 & 1124.1  &  2.7 & 92.6 & -4.995 &  0.011 \\
 6701.572178 & 1069.7  &  3.2 & 76.1 & -4.993 &  0.016 \\
 6702.578027 & 1044.5  &  3.6 & 83.2 & -4.967 &  0.015 \\
 6728.651408 & 1105.1  &  3.8 & 91.5 & -5.020 &  0.021  \\
 6786.609623 & 1143.6  &  4.6 & 103.4 & -4.952 &  0.027  \\
 6787.608024 & 1112.3  &  3.9 & 101.6 & -4.983 &  0.023 \\
 6817.482897 & 1176.6  &  2.7 & 83.9 & -5.022 &  0.013  \\
 6819.529437 & 1115.1  &  3.6 & 78.6 & -4.987 &  0.019 \\
 6820.503665 & 1101.6  &  5.3 & 78.4 & -5.035 &  0.040  \\
 6861.396206 & 1150.9  &  2.5 & 87.9 & -5.001 &  0.011  \\
 6874.367191 & 1144.3  &  3.7 & 94.1 & -4.996 &  0.020  \\
 6875.369909 & 1119.5  &  9.1 & 67.0 & -4.975 &  0.076 \\
 6987.775868 & 1204.0  &  3.6 & 86.8 & -4.968 &  0.019  \\
 6998.737459 & 1198.9  &  5.5 & 81.9 & -4.950 &  0.032 \\
 6999.772079 & 1166.2  &  5.7 & 107.6 & -5.021 &  0.044  \\
 7005.756862 & 1234.1  &  9.3 & 71.0 & -5.034 &  0.086  \\
 7011.773901 & 1226.4  &  2.7 & 90.9 & -4.997 &  0.012  \\
 7028.700777 & 1240.1  &  3.5 & 101.0 & -5.015 &  0.019  \\
 7029.691968 & 1216.8  &  3.5 & 87.4 & -5.001 &  0.018  \\
 7050.675470 & 1224.3  &  4.8 & 84.4 & -4.985 &  0.030  \\
 7068.780530 & 1218.8  &  4.7 & 109.7 & -4.971 &  0.028  \\
 7109.622389 & 1170.3  &  5.0 & 87.8 & -5.000 &  0.031  \\
 7112.604654 & 1155.1  &  8.0 & 78.7 & -4.996 &  0.069  \\
 7113.520997 & 1209.9  &  3.7 & 86.7 & -4.977 &  0.021  \\
 7114.443438 & 1212.0  &  5.0 & 74.6 & -4.960 &  0.032 \\
 7115.451479 & 1219.3  &  2.6 & 90.4 & -5.003 &  0.012  \\
 7124.384992 & 1203.2  &  4.1 & 89.8 & -4.958 &  0.024 \\
 7124.395594 & 1203.2  &  3.7 & 95.3 & -4.983 &  0.022 \\
 7124.406149 & 1206.1  &  3.3 & 101.8 & -4.989 &  0.018 \\
 7124.416785 & 1217.2  &  3.2 & 109.7 & -5.013 &  0.019 \\
 7124.427595 & 1213.6  &  3.4 & 97.6 & -4.994 &  0.020 \\
 7124.438266 & 1210.4  &  3.3 & 85.0 & -4.992 &  0.018 \\
 7124.448937 & 1215.6  &  3.4 & 78.9 & -4.968 &  0.018 \\
 7124.459481 & 1212.7  &  3.4 & 99.1 & -5.013 &  0.020 \\
 7124.470175 & 1214.5  &  3.6 & 86.5 & -5.006 &  0.022 \\
 7124.481309 & 1198.8  &  3.0 & 82.7 & -5.013 &  0.016 \\
 7124.491679 & 1202.4  &  2.9 & 94.5 & -4.994 &  0.014 \\
 7124.502581 & 1198.6  &  3.1 & 80.4 & -5.001 &  0.016 \\
 7124.513125 & 1196.5  &  3.2 & 85.3 & -4.987 &  0.016 \\
 7124.524155 & 1186.0  &  3.1 & 83.6 & -5.006 &  0.016 \\
 7124.534618 & 1187.8  &  3.3 & 84.4 & -4.988 &  0.018 \\
 7124.543668 & 1173.8  &  7.7 & 97.4 & -4.956 &  0.056 \\
 7124.556156 & 1179.0  &  3.2 & 91.7 & -5.013 &  0.018 \\
 7124.566665 & 1180.4  &  3.1 & 98.1 & -5.005 &  0.017 \\
 7124.577441 & 1171.2  &  3.1 & 92.9 & -5.015 &  0.017 \\
 7124.587949 & 1176.1  &  3.2 & 89.6 & -4.997 &  0.017 \\
 7124.599130 & 1174.4  &  3.3 & 72.1 & -5.001 &  0.019 \\
 7124.609685 & 1170.2  &  3.2 & 91.4 & -5.011 &  0.018 \\
 7124.620229 & 1182.9  &  3.1 & 103.7 & -4.998 &  0.017 \\
 7124.631085 & 1181.3  &  3.4 & 83.3 & -4.996 &  0.018 \\
 7124.641640 & 1183.5  &  3.4 & 69.9 & -5.002 &  0.018 \\
 7124.652496 & 1189.0  &  3.4 & 91.8 & -4.976 &  0.017 \\
 7124.663202 & 1176.4  &  3.5 & 81.3 & -4.982 &  0.018 \\
 7124.673908 & 1188.7  &  3.3 & 83.9 & -4.985 &  0.016 \\
 7124.684347 & 1186.2  &  3.2 & 89.2 & -4.987 &  0.017 \\
 7124.695354 & 1182.3  &  2.9 & 101.9 & -4.995 &  0.015 \\
 7124.705932 & 1179.9  &  2.9 & 101.8 & -4.980 &  0.015 \\
 7142.537170 & 1144.0  &  3.6 & 89.5 & -4.990 &  0.021  \\
 7143.470202 & 1158.9  &  3.4 & 86.3 & -5.018 &  0.020  \\
 7144.467594 & 1196.6  &  3.8 & 84.4 & -5.003 &  0.023  \\ 
 7150.499394 & 1155.1  &  4.6 & 90.2 & -5.004 &  0.026  \\
 7159.414630 & 1164.7  &  3.2 & 99.1 & -5.024 &  0.019  \\
 7170.469914 & 1214.6  &  2.7 & 83.9 & -5.010 &  0.012  \\
 7173.479806 & 1143.8  &  5.2 & 112.2 & -4.961 &  0.033  \\
 7174.499116 & 1140.4  &  5.3 & 91.6 & -4.937 &  0.032  \\
 7175.462458 & 1179.1  &  2.7 & 99.2 & -4.996  & 0.013 \\
 7176.476235 & 1212.3  &  3.8 & 82.5 & -4.994  & 0.021  \\ 
 7205.414495 & 1136.2  &  2.5 & 90.2 & -5.000   & 0.011  \\
 7206.421521 & 1157.3  &  3.1 & 80.1 & -4.968   & 0.014  \\
 7207.432852 & 1189.7  &  3.8 & 87.3 & -4.952   & 0.021  \\
\noalign{\smallskip}
\hline
\noalign{\smallskip}
\noalign{\smallskip}
\hline
 \end{tabular} 
 \end{table}
}

\onltab{
\begin{table}
    \caption{List of the new relative radial velocities of KELT-6 measured with the TRES spectrograph and used for the first time in this work.}
    \label{table:radveltres}
        \centering
        \scriptsize
        \begin{tabular}{ccc}
        \hline
        \noalign{\smallskip}
        \noalign{\smallskip}
        \hline
    \noalign{\smallskip}
    Time  & RV & RV error \\  
        (BJD$_{UTC}$ - 2,450,000) & (m s$^{-1}$) & (m s$^{-1}$)  \\ 
 6640.021901 & -30.50  &   36.12 \\
 6643.998845 &  35.66  &   23.47 \\
 6698.953413 &  86.48  &   24.29 \\
 6702.883586 & -15.28  &   28.33 \\
 6729.914621 &  64.77  &   24.93 \\
 6733.902128 &   4.10  &   19.52 \\
 6801.709729 &  59.28  &   22.98 \\
 6803.711379 & -15.48  &   12.28 \\
 6804.800386 & -14.19  &   22.87 \\
 6812.690753 &  54.72  &   29.16 \\
 6816.685639 &  73.70  &   18.88 \\
 6824.685293 &  94.65  &   14.15 \\
 6993.029954 & 107.42  &   22.14 \\
 7055.946535 &  94.92  &   18.22 \\
 7090.798944 & 172.10  &   19.99 \\
 7094.908058 & 101.59  &   19.32 \\
 7110.916771 &  73.25  &   20.13 \\
 7121.849369 & 133.02  &   17.07 \\
 7141.862950 & 35.84   &   17.49 \\
 7145.767451 & 156.82  &   17.82 \\
 7149.696112 & 42.03   &   18.32 \\
 7169.684005 & 133.14  &   18.14 \\
\noalign{\smallskip}
\hline
\noalign{\smallskip}
\noalign{\smallskip}
\hline
 \end{tabular} 
 \end{table}
} 
        

    \begin{table}  
    \caption{Orbital and physical parameters for KELT-6\,b and KELT-6\,c.
    }
    \label{table:orbitalparam}
        \centering
        \begin{scriptsize}
        \begin{tabular}{ccc}
        \hline
    \noalign{\smallskip}
    Parameter &  KELT-6\,b & KELT-6\,c \\ 
    \noalign{\smallskip}
    \hline
    \noalign{\smallskip}
     \textbf{Primary transit analysis} & &  \\ [2pt]
        $T_c$ [BJD$_{\rm TDB}-2,450,000$] & $7124.50954\pm5.7 \cdot 10^{-4}$ & \\ [0.5pt] 
        $P$ [days] & $7.845582\pm7 \cdot 10^{-6}$ & \\ [0.5pt] 
        ($R_{\rm p}/R_{\rm s}$)\tablefootmark{a}  & $0.077613_{-0.0009}^{+0.0010} $ &  \\ [0.5pt]
        $\textit{b}$\,\tablefootmark{a,b}  & $0.20_{-0.13}^{+0.14} $ & \\[0.5pt] 
        $\textit{i}$\,\tablefootmark{a,c} [deg]  & $88.81_{-0.91}^{+0.79} $ & \\[0.5pt] 
     \textbf{RV analysis} & & \\ [2pt]
        $P$\,\tablefootmark{d} [days] &   $7.8455821\pm7 \cdot 10^{-6}$ & $1276_{-67}^{+81}$ \\[0.5pt]
        $K$\,\tablefootmark{d} [m s$^{-1}$] & $41.8\pm1.1 $   & $65.7_{-2.4}^{+2.6}$ \\[0.5pt] 
        $e$\, \tablefootmark{e} & $ 0.029_{-0.013}^{+0.016} $  &  $0.21_{-0.036}^{+0.039}$ \\ [0.5pt]        
        ($\sqrt{e} \sin \omega$)\tablefootmark{d} & $-$0.083$_{-0.086}^{+0.120}$  & $-0.454\pm0.042$ \\ [0.5pt]
        ($\sqrt{e} \cos \omega$)\tablefootmark{d} & $0.126_{-0.046}^{+0.035}$     &  $-0.011\pm0.071$ \\[0.5pt] 
        $\omega$\,\tablefootmark{e}  [deg]  & $308_{-272}^{+30}$  &  $  268.7\pm8.8$ \\ [0.5pt] 
        $T_c$\,\tablefootmark{d} [$\rm BJD_{TDB}-2,450,000$]  &  $7124.50954\pm5.8 \cdot 10^{-4} $  & $7432_{-32}^{+39}$ \\ [0.5pt]
        $\gamma_{\rm TRES}$\tablefootmark{d} $[\ms]$ & \multicolumn{2}{c}{ $46.3\pm3.8$}\\ [0.5pt]
        $\gamma_{\rm HIRES}$\tablefootmark{d} $[\ms]$ & \multicolumn{2}{c}{ $22.1\pm8.5$}\\[0.5pt]
        $\gamma_{\rm HARPS-N}$\tablefootmark{d} $[\ms]$  & \multicolumn{2}{c}{ $1128.9\pm3.0$} \\[0.5pt]
        ($jitter_{\rm TRES}$)\, \tablefootmark{d}  $[\ms]$ & \multicolumn{2}{c}{  $ 14.2 \pm 4.5$ }\\[0.5pt] 
        ($jitter_{\rm HIRES}$)\, \tablefootmark{d} $[\ms]$ & \multicolumn{2}{c}{  $ 8.1_{-1.8}^{+2.3}$ } \\[0.5pt] 
        ($jitter_{\rm HARPS-N}$)\,\tablefootmark{d} $[\ms]$ & \multicolumn{2}{c}{  $ 2.2 \pm 1.2$ }\\[0.5pt]
         ($M_{\rm p}\sin i$)\,\tablefootmark{e,f}  $~[\Mjup]$  &   $0.442\pm0.019$ &  $3.71\pm0.21$  \\[0.5pt]
         $R_{p}$\,\tablefootmark{g} $ ~[\Rjup]$  & $1.18\pm0.11$ & \\[0.5pt] 
         $a_{p}$\tablefootmark{h} [AU] &  0.080$\pm$0.001 & $2.39\pm0.11$ \\ [0.5pt]
         $\gamma_{RM}$\tablefootmark{i}  $[\ms]$  & 1190.1 $\pm$ 1.4 & \\[0.5pt] 
         $\lambda$\,\tablefootmark{i} [deg] & $-$36$\pm$11 & \\[0.5pt] 
        \hline
        \end{tabular} 

        \tablefoot{\scriptsize
        \tablefoottext{a}{Co14}
        \tablefoottext{b}{Impact parameter.}
        \tablefoottext{c}{Inclination of the orbital plane.}
        \tablefoottext{d}{Fitted.}
        \tablefoottext{e}{Derived.}
        \tablefoottext{f}{For KELT-6\,b this corresponds to the real mass of the planet, derived by assuming the best estimate of the inclination angle of the orbital plane provided by Co14.}
         \tablefoottext{g}{Derived from the transit parameter $R_{\rm p}/R_{\rm s}$ in Co14 and our value of $R_{\rm s}$ through a Monte-Carlo analysis.}
         \tablefoottext{h}{Derived from the third Kepler's law through a Monte-Carlo analysis, using our updated values for the stellar mass and planetary orbital periods.}
         \tablefoottext{i}{Fitted (Rossiter-McLaughlin effect).}}
             \end{scriptsize}
    \end{table}

\subsection{The Rossiter-McLaughlin effect for KELT-6\,b}
\label{sec:rml}
\begin{figure}
 \centering
 \includegraphics[width=6cm]{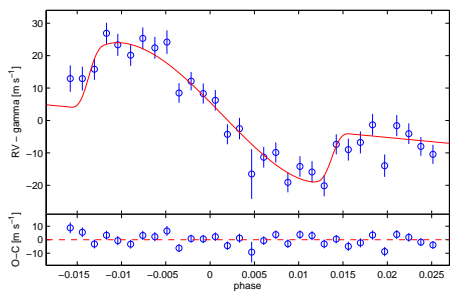}
  \caption{HARPS-N RV time series spanning the transit of KELT-6\,b occurred on 2015 April 11 (open blue circles) and
  our best-fit model of the Rossiter-McLaughlin effect (red solid line).}
  \label{Fig:RM-Kelt6b}
\end{figure}

    The RV time series covering the transit of KELT-6\,b was analysed using the numerical model and the least-squares
fitting algorithm described in \citet{covino13} and \cite{esposito14}.
Four parameters were set free\footnote{$\gamma$: barycentric RV at mid-transit epoch; $\lambda$: projected angle between the planetary orbital axis and the stellar spin axis; $V\sin I_*$: projected stellar rotational velocity;
$\mu$: limb-darkening coefficient of a linear law.}, while all other relevant parameters were kept fixed to the values derived
from the spectroscopic and photometric analyses.
The best-fit values are reported in Tables \ref{table:stellarparam} and \ref{table:orbitalparam}, together with their uncertainties derived by means of a bootstrapping method. The best-fit RM model is shown in Fig. \ref{Fig:RM-Kelt6b}. Results show that KELT-6\,b moves in a prograde orbit and is slightly misaligned with respect to the stellar spin axis with a projected spin-orbit angle $\lambda$=$-$36$^{\circ}\pm11^{\circ}$. The lack of enough RV data in the pre-transit phase means that a value of $\lambda$ closer to zero cannot be completely ruled out, but our solution comes with a lower $\chi^{2}$ than for the case of $\lambda$ fixed to 0 (1.39 $vs.$ 1.89).     
    
\section{Discussion}
\label{sec:discussion}

KELT-6 joins a small number of host stars with a transiting planet and a measured RM effect for which there is 
evidence of outer companions. To our knowledge they are HAT-P-13, HAT-P-17, and WASP-8 \citep{knutson14}, the last two hosting a transiting planet with the orbital period close to that of KELT-6b (respectively, P=10.3 and 8.2 days) and with WASP-8 being misaligned ($\lambda$=$-123.3_{-4.4}^{+3.4}$)\footnote{see http://www.astro.keele.ac.uk/jkt/tepcat/rossiter.html}.
Our results allow the description of the KELT-6 system architecture in some detail, and considerations can be made about its dynamical evolution and stability. 

We tested the coplanar high-eccentricity migration hypothesis (CHEM, \citealt{petrovich15a}) as a possible mechanism to explain the current orbital elements of KELT-6. CHEM explores the possibility that hot Jupiters are formed through secular gravitational interactions with an outer planet on an initial eccentric orbit, and then circularized by tides with the host star at periastron. The mechanism describes the planetary migration as occurring on nearly the same plane where the planets formed, assuming a low mutual inclination of the orbital planes ($\lesssim20^{\circ}$), and it predicts that hot Jupiters should have distant and more massive companions with moderately high eccentricities ($e\sim$ 0.2-0.5). 
CHEM reproduces the observed architecture of KELT-6, provided that the inner planet, initially with \textit{e}$\gtrsim$0.5, started migration inside $\sim$1 au, and there was an initial moderate mutual inclination of $\sim$10$^{\circ}$-20$^{\circ}$ to account for the measured angle $\lambda$. This suggests that the current mutual inclination of the orbital planes could be lying in this range. We note that the system should be old enough for the orbit of KELT-6b to have been circularized. In fact, the circularization time with the current system parameters is about 0.45 Gyr, assuming a modified tidal quality factor $Q^{\prime}_{\rm p}=10^{5}$ for the planet \citep{lainey09}, where most of the tidal dissipation is expected to have occurred in the case of an initially eccentric orbit (see \citealt{jackson08}). 
We speculate that the initial conditions required by CHEM to explain this system could be the result of a preceding phase of planet-planet scattering. A system with more than two planets in nearly circular and coplanar orbits at $\gtrsim$1 au might have become unstable, losing planets by ejections and leading to eccentricity excitation and moderate change (by a factor up to $\sim$2) in the semi-major axis of the inner planet \citep{juric08,chatter08}.

We investigated the dynamical stability of the KELT-6 system by using the empirical relation (17) in \cite{petrovich15b}, who studied the final results of the evolution on long timescales of two-planet systems with arbitrary eccentricities and mutual inclinations against either ejections or collisions with the host star. We found that the relation is satisfied well, implying a stability preserved on a timescale of a few Myr (i.e. the range of validity of Petrovich's formula), as expected for a system that is already $\sim$5 Gyr old.


\begin{acknowledgements}
GAPS acknowledges support from INAF through the “Progetti Premiali” funding scheme of the Italian Ministry of Education, University, and Research. MD acknowledges support from INAF-OATo through grant $\#$35/2014, and thanks the Astronomical Observatory of the Autonomous Region of the Aosta Valley for its support with computing infrastructure. IRAF is distributed by the National Optical Astronomy Observatories, operated by the Association of Universities for Research in Astronomy, Inc., under cooperative agreement with the National Science Foundation. We made use of the SIMBAD database and VizieR catalogue access tool, operated at the CDS, Strasbourg, France.
\end{acknowledgements}


\bibliographystyle{aa} 
\bibliography{ref.bib} 

\end{document}